\documentclass[pra,nofootinbib,showpacs,showkeys,draft]{revtex4}

\def\mb#1         {\mbox{\boldmath $#1$}}

\begin{document}
%
%
\title{Spinor Operator Giving Both Angular Momentum and Parity}
%
%
\author{Takayuki Matsuki}
\email[E-mail: ]{matsuki@tokyo-kasei.ac.jp}
\affiliation{Tokyo Kasei University,
1-18-1 Kaga, Itabashi, Tokyo 173, JAPAN}
\author{Kentarou Mawatari}
\email[E-mail: ]{mawatari@radix.h.kobe-u.ac.jp}
\author{Toshiyuki Morii}
\email[E-mail: ]{morii@kobe-u.ac.jp}
\affiliation{Graduate School of Science and Technology,
Kobe University,\\ Nada, Kobe 657-8501, JAPAN}
\author{Kazutaka Sudoh}
\email[E-mail: ]{sudou@rarfaxp.riken.go.jp}
\affiliation{Radiation Laboratory, RIKEN 
(The Institute of Physical and Chemical Research), \\ 
Wako, Saitama 351-0198, JAPAN}

\date{August 28, 2004}

\begin{abstract}
In heavy quark effective theory, heavy mesons which contain a heavy
quark (or antiquark) are classified by $s_\ell^{\,\pi_\ell}$, i.e., 
the total angular momentum  $s_\ell$ and the parity $\pi_\ell$ of the
 light quark degrees of freedom around a static heavy quark. In this
 case,  however, one needs to separately estimate the parity other than
 the angular momentum of a light quark to describe heavy mesons.

A new operator $K$ was proposed some time ago by two of us 
(T.M. and T.M.). In this Letter, we show that the quantum number $k$
of this operator is enough to describe both the total angular
momentum of the light quark degrees of freedom and the parity of a heavy
 meson, and derive a simple relation between $k$ and $s_\ell^{\,\pi_\ell}$.
\end{abstract}

\medskip
\preprint{}
\maketitle

Recent discovery of narrow meson states $D_{sJ}(2317)$ and $D_{sJ}(2460)$
by BaBar and the following confirmation by CLEO and Belle \cite{Dsj} has 
triggered a series of study on spectroscopy of heavy mesons again.
Though $D_{sJ}(2317)$ and $D_{sJ}(2460)$ are assigned as $j^P=0^+$ and
$1^+$, respectively,
their masses are significantly smaller than the predictions based on many
of potential models \cite{GI}. 
To explain these masses, Bardeen, Eichten, Hill and others \cite{BEH, BH} 
proposed an interesting idea of an effective Lagrangian with chiral 
symmetries of light quarks and heavy quark symmetry.
The heavy meson states with the total angular momentum $j=0$ and $j=1$
related to $s_{\ell}$ (the total angular momentum of the light quark degrees of
freedom) $=1/2$ make the parity doublets $(0^-, 0^+)$ and $(1^-, 1^+)$,
respectively, and the members in these doublets degenerate in the 
limit of chiral symmetry.
Furthermore, the two states $(0^-, 1^-)$ degenerate in the 
limit of heavy quark symmetry, as well as $(0^+, 1^+)$. 
These doublets are called the heavy spin multiplets.

These newly discovered states are well classified in heavy quark
effective theory, i.e., in terms of $s_\ell^{\,\pi_\ell}$, 
where $s_\ell$ and $\pi_\ell$
represent the total angular momentum and the parity of the light quark
degrees of freedom around a static heavy quark, respectively. In this
case, however, one has to separately estimate the parity and the angular
momentum of a light quark for each heavy meson state.

Some time ago, two of the authors (T.M. and T.M.) proposed 
a new bound state equation for atomlike mesons, i.e., heavy mesons
composed of a heavy quark and a light antiquark, and they also proposed
a new operator $K$ which can classify heavy mesons well \cite{MM}. 
In this Letter, we show that this operator $K$, given by
Eq. (\ref{k_quantum}) below, has the information about not only $s_\ell$
but also the parity of heavy mesons and naturally explains the heavy
spin multiplets. That is, only the quantum number $k$ corresponding to
the operator $K$ can reproduce both the total angular momentum of the
light quark degrees of freedom {\it and} the parity of a heavy meson. We
also discuss the relation between $k$ and $s_\ell^{\,\pi_\ell}$. \\

Let us consider a heavy meson composed of a heavy quark
$Q$ and a light antiquark $\bar q$.
The effective Hamiltonian of this system is obtained by applying
the Foldy-Wouthuysen-Tani (FWT) transformation to the heavy quark
$Q$.  One can formulate the equation so that we can cast the 
structure of the eigenvalue equation into a simple form and make 
the Dirac-like equation in the large limit of the heavy quark mass $m_Q$ 
\cite{MM}. 
In order to show why we can introduce a new operator
$K$ for heavy mesons, we consider the equation with 
$1/m_Q$ corrections neglected, whose contribution should be important in 
numerical analysis of spectroscopy. 
 
The lowest energy for the $Q\bar q$ bound state is given by $m_Q+E_0^a$
after solving the equation \cite{MM} 
\begin{equation}
  H_0 \otimes\psi _0^a = E_0^a\psi _0^a, \qquad
  H_0 =
  \vec\alpha_q\cdot\vec p_q+\beta_q\left(m_q+S(r)\right)+V(r), \label{0th}
\end{equation}
where $a$ expresses all the quantum numbers and quantities with the 
subscript $q$ mean those for a light antiquark.  $S(r)$ is a confining scalar
potential and $V(r)$ is a Coulombic vector potential at short distances. Both
potentials have dependence only on $r$, the relative distance between $Q$ and
$\bar q$. With a symbol $\otimes$, one should note that gamma
matrices for a light antiquark be multiplied from left with the wave function
while those for a heavy quark from right.

Using the $2 \times 2$ matrix eigenfunctions $y_{j\,m}^k$ of angular part 
defined below and the radial functions $f_k$ and $g_k$, the
$4\times 4$ matrix solution to Eq. (\ref{0th}) is given by \cite{MM}
\begin{eqnarray}
  \psi _0^a &=& \left( {\matrix{ 0  & \Psi _{j\,m}^k(\vec r) }} \right),
  \label{0thsols1}
  \\
  \Psi _{j\,m}^k(\vec r) &=& \frac{1}{r}
  \left( {\matrix{f_k(r)\;y_{j\,m}^k\cr ig_k(r)\;y_{j\,m}^{-k}\cr}} \right),
  \label{0thsols2}
\end{eqnarray}
where $j$ and $m$ are the total angular momentum of a heavy meson and its
$z$-component, respectively. The total angular momentum of a heavy
meson is the sum of the total angular momentum of the light quark
degrees of freedom $\vec S_\ell$ and the heavy quark spin  
${1\over 2}\vec \Sigma_Q$:  
\begin{equation}
  {\vec J} = {\vec S_\ell} +{1\over 2}\vec \Sigma_Q\qquad {\rm with} \quad
  {\vec S_\ell}=\vec L + {1\over 2}\vec \Sigma_q,
\end{equation}
where ${1\over 2}\vec\Sigma_q\ (={1\over 2}\vec\sigma_q\;1_{2\times 2})$ 
and $\vec L$ are
the 4-component spin and the orbital angular momentum of a light antiquark, 
respectively.
Furthermore, $k$ is the quantum number of the spinor operator $K$, which
was introduced in Eq. (20) of Ref. \cite{MM}, defined by
\begin{equation}
  K = -\beta_q \left( \vec \Sigma_q \cdot \vec L + 1 \right),
  \qquad
  K\, \Psi_{j\,m}^k = k\, \Psi_{j\,m}^k.
  \label{k_quantum}
\end{equation}
It is interesting to note that the same form of
the operator $K$ is defined in the case of a single Dirac particle in a
central potential \cite{JJ}. It is remarkable that in our approach $K$
can be defined even for a heavy meson which is a two-body bound system
composed of a heavy quark and a light antiquark.

Here we show that there is a relation between $k$ and $s_\ell$, being
often used in heavy quark effective theory. Let us calculate the square of $K$.
\begin{eqnarray}
  K^2 
  &=&
  \left( \Sigma_q \right)_i \left( \Sigma_q \right)_j L_i L_j
  + 2 \vec \Sigma_q \cdot \vec L + 1
  =
  {\vec L}^2 + \vec \Sigma_q \cdot \vec L + 1
  \nonumber \\
  &=&
  {\vec S_\ell}^2 + \frac{1}{4}.
\end{eqnarray}
Therefore, the operator $K^2$ is equivalent to ${\vec S_\ell}^2$ and it holds
\begin{equation}
  k = \pm \left( s_\ell + \frac{1}{2} \right) \quad {\rm or} \quad 
s_\ell = \left| k \right| - \frac{1}{2}. \label{k}
\end{equation}

Now, let us briefly summarize the properties of the  
eigenfunctions $y_{j\,m}^k$, whose details are given in \cite{MM}. 
To begin with, we need to introduce the so-called vector
spherical harmonics which are defined by \cite{landau}
\begin{eqnarray}
  \vec Y_{j\,m}^{(\rm L)} = -\vec n\,Y_j^m, \quad
  \vec Y_{j\,m}^{(\rm E)} = {r \over {\sqrt {j(j+1)}}}\vec \nabla Y_j^m,
  \quad
  \vec Y_{j\,m}^{(\rm M)} = -i\vec n\times \vec Y_{j\,m}^{(\rm E)},
\end{eqnarray}
where $Y_j^m$ are the spherical polynomials and $\vec n=\vec r/r$. 
These vector spherical harmonics are nothing but a set of eigenfunctions
for a spin-1 particle. 
$\vec Y_{j\,m}^{(\rm A)}$ (A=L, M, E) are eigenfunctions of 
${\vec J}^2$ and $J_z$, having the eigenvalues $j(j+1)$ and $m$. 
The parities are assigned as
$(-)^{j+1}$, $(-)^j$, $(-)^{j+1}$ for A=L, M, E, respectively, since 
$Y_j^m$ has a parity $(-)^j$.

In order to diagonalize the leading Hamiltonian of Eq. (\ref{0th}) 
in the $k$ space, it is necessary to make $\vec Y_{j\,m}^{(\rm A)}$ and
$Y_j^m$ into the spinor representation 
$y^k_{j\,m}$ by the following unitary transformation
\begin{equation}
  \left( {\matrix{{y_{j\,m}^{-(j+1)}}\cr {y_{j\,m}^j}\cr}} \right)
  =U\left( {\matrix{{Y_j^m}\cr 
  {\vec \sigma \cdot \vec Y_{j\,m}^{(\rm M)}}\cr}} \right),
  \qquad\left( {\matrix{{y_{j\,m}^{j+1}}\cr {y_{j\,m}^{-j}}\cr}} \right)
  =U\left( {\matrix{{\vec \sigma \cdot \vec Y_{j\,m}^{(\rm L)}}\cr
  {\vec \sigma \cdot \vec Y_{j\,m}^{(\rm E)}}\cr}} \right),
\end{equation}
where the orthogonal matrix $U$ is introduced as
\begin{equation}
  U={1 \over {\sqrt {2j+1}}}\left( {\matrix{{\sqrt {j+1}}&{\sqrt j}\cr
  {-\sqrt j}&{\sqrt {j+1}}\cr}} \right). \label{eq:app:u}
\end{equation}
$y_{j\,m}^k$ are $2 \times 2$ matrix
eigenfunctions of three operators, $\vec J^{\,2}$,
$J_z$, and $\vec \sigma_q \cdot \vec L$ with eigenvalues, $j(j+1)$,
$m$, and $-(k+1)$, respectively, and have the interesting properties
\begin{eqnarray}
  \left(\vec \sigma_q \cdot \vec n \right) \otimes y_{j\,m}^k
  &=& - y_{j\,m}^{-k}, \label{k_eigen1} \\
  (\vec \sigma_q \cdot \vec L ) \otimes y_{j\,m}^k
  &=& - (k+1) \;y_{j\,m}^k, \label{k_eigen2}
\end{eqnarray}
where the quantum number $k$ can take only values as shown in Eq. 
(\ref{k})
\begin{equation}
 k=\pm j \quad {\rm or}\quad \pm (j+1). \label{kj}
\end{equation}
It should be noted that $k$ is nonzero since $\vec Y_{0\;0}^{(\rm M)}$
does not exist.

Substituting Eqs. (\ref{0thsols1}) and (\ref{0thsols2}) 
into Eq. (\ref{0th}) and using Eqs. (\ref{k_eigen1}) and (\ref{k_eigen2}), 
one can eliminate the angular part of the wave 
function, $y_{j\,m}^k$, and obtain the radial equation as follows,
\begin{equation}
  \left( {\matrix{{m_q+S+V}&{-\partial _r+{k \over r}}\cr
  {\partial _r+{k \over r}}&{-m_q-S+V}\cr
  }} \right) \Psi _k(r) = E^k_0\;\Psi _k(r), 
  \qquad
  \Psi _k(r)\equiv\left( {\matrix{{f_k\left( r \right)}\cr
  {g_k\left( r \right)}\cr}} \right),
\end{equation}
which depends on $k$ alone.  This is quite similar to a one-body Dirac
equation in a central potential.
Since $K$ commutes with $H_0$ and 
the states $\Psi _{j\,m}^k$ have the same energy $E_0^k$ with
different values of $j$, these states are degenerate with the
same value of $k$ at the lowest order in $1/m_Q$.

The parity $P'$ of the eigenfunction $\psi_0^a$ is determined by the upper
(``large'') two-by-two components $y_{j\,m}^k$ as
\begin{equation}
 P'= \left\{
  \begin{array}{lccclcl}
   (-1)^j     &\quad & {\rm for} &\quad 
        & \Psi^{-(j+1)}_{j\,m} & {\rm and} & \Psi^j_{j\,m}\\
   (-1)^{j+1} &\quad & {\rm for} &\quad 
        & \Psi^{j+1}_{j\,m} & {\rm and} & \Psi^{-j}_{j\,m}
  \end{array}
 \right. \label{ppsi}
\end{equation}
Thus, using the relations of Eqs. (\ref{kj}) and (\ref{ppsi})  
and taking into account the
intrinsic parity of the light antiquark, one can simply write the parity
of a heavy meson as 
\begin{equation}
  P = -P'=\frac{k}{|k|} (-1)^{|k| + 1}
\label{parity}
\end{equation}
Notice that the parity $P$ of the whole system is equal to 
the parity $\pi_\ell$ of the light quark degrees of freedom, as can be
seen in TABLE \ref{table}, since the
intrinsic  parity of a heavy quark is $+1$.

In heavy quark effective theory, heavy mesons are normally classified
in terms of $s_\ell^{\pi_\ell}$, since at the lowest order
heavy quarks in those mesons are considered to be static, namely it
stays rest at the center of a heavy meson system. 
In this work, we have found that (i) the parity of a heavy meson and 
(ii) the total angular momentum of the light quark degrees of 
freedom can be reproduced in terms of $k$ alone as seen 
from Eqs. (\ref{parity}) and (\ref{k}), respectively.
We have also found that the degeneracy between members in each heavy spin
multiplet, $(0^-,\;1^-)$ and $(0^+,\;1^+)$,
is automatic in our
approach \cite{MM}, while the method using the effective Lagrangian with
heavy-quark as well as chiral symmetries must force degeneracy among parity
doublets to construct such a Lagrangian \cite{BEH, BH}.
These are {\it the main results of this paper}. \\ 
\begin{table*}[t]
\caption{States classified by various quantum numbers}
\label{table}
\begin{tabular*}{13cm}{c|@{\extracolsep{\fill}}ccccccc}
\hline
\hline
  \makebox[1.7cm]{$j^P$}   &   $0^-$   &   $1^-$   &   $0^+$   &   
  $1^+$   &   $1^+$   &   $2^+$   &   $1^-$   \\
  $k$   &   -1   &   -1   &   1   &   
   1    &   -2   &   -2   &   2   \\
  $s_\ell^{\pi_\ell}$ & ${1\over2}^-$ & ${1\over2}^-$ & ${1\over2}^+$ & 
  ${1\over2}^+$ & ${3\over2}^+$ & ${3\over2}^+$ & ${3\over2}^-$ 
  \\
  $^{2s+1}l_j$ & $^1S_0$ & $^3S_1$  & $^3P_0$  & $^3P_1$, $^1P_1$ &
   $^1P_1$, $^3P_1$  & $^3P_2$  & $^3D_1$  
  \\ \hline 
  $\Psi_j^k$ & $\Psi_0^{-1}$ & $\Psi_1^{-1}$ & $\Psi_0^1$ & $\Psi_1^1$ &
  $\Psi_1^{-2}$ & $\Psi_2^{-2}$ & $\Psi_1^2$ \\
\hline
\hline
\end{tabular*}
\end{table*}

As our summary, several states are classified by various quantum numbers
in TABLE \ref{table}.
The states with different $j$ but with the same parity $P$ make a 
heavy spin multiplet of heavy mesons, which corresponds to
{\it heavy quark symmetry} in heavy quark effective theory.
One can see that $k$ naturally explains the heavy spin doublets.

Before closing our discussions, we comment about $k$ from the
phenomenological point of view.
The lowest order solution satisfies degeneracy in $k$ since the energy depends
only on $k$, i.e., $j^P=0^-$ and $1^-$ states have the same mass, so are the
$0^+$ and $1^+$ states. This degeneracy is resolved by
including higher order terms in $1/m_Q$ \cite{MM} and one can 
phenomenologically discuss mass spectra of
these heavy mesons even though some objections \cite{BCL} 
for using a potential model exist.  A comprehensive analysis on mass
spectra of heavy mesons including $D_{sJ}(2317)$ and $D_{sJ}(2460)$
is in progress.

\def\Journal#1#2#3#4{{#1} {\bf #2}, #3 (#4)}
\def\NIM{Nucl. Instrum. Methods}
\def\NIMA{Nucl. Instrum. Methods A}
\def\NPB{Nucl. Phys. B}
\def\PLB{Phys. Lett. B}
\def\PRL{Phys. Rev. Lett.}
\def\PRD{Phys. Rev. D}
\def\ZPC{Z. Phys. C}
\def\EPJ{Eur. Phys. J. C}
\def\PR{Phys. Rept.}
\def\IJM{Int. J. Mod. Phys. A}


\begin{thebibliography}{00}
\newcommand{\etal}{{\it et al.}}
\bibitem{Dsj} BaBar Collaboration, B. Aubert \etal, 
                \Journal{\PRL}{90}{242001}{2003};
              CLEO Collaboration, D. Besson \etal, 
               \Journal{\PRD}{68}{032002}{2003}; 
              Belle Collaboration, P. Krokovny \etal,
	       \Journal{\PRL}{91}{262002}{2003};
	      Y. Mikami \etal, 
               {\it ibid.} {\bf 92}, 012002 (2004).
\bibitem{GI} See, for example, S. Godfrey and N. Isgur, 
              \Journal{\PRD}{32}{189}{1985};
               S. Godfrey and R. Kokoski, 
               {\it ibid.} {\bf 43}, 1679 (1991). 
\bibitem{BEH} W. A. Bardeen, E. J. Eichten, and C. T. Hill, 
              \Journal{\PRD}{68}{054024}{2003}.
\bibitem{BH} W. A. Bardeen and C. T. Hill, 
             \Journal{\PRD}{49}{409}{1994}; 
             M. A. Nowak, M. Rho, and I. Zahed, 
             {\it ibid.} {\bf 48}, 4370 (1993); 
             A. Deandrea, N. Di Barolomeo, R. Gatto, G. Nrdulli, 
             and A. D. Plosa, 
             {\it ibid.} {\bf 58}, 034004 (1998); 
             A. Hiorth and J. O. Eeg, 
             {\it ibid.} {\bf 66}, 074001 (2002).
\bibitem{MM} T. Matsuki and T. Morii, 
             \Journal{\PRD}{56}{5646}{1997}.
\bibitem{JJ} M. E. Rose, {\it Relativistic Electron Theory}, John Wiley
	\& Sons, 1961; 
             J. J. Sakurai, 
             {\it Advanced Quantum Mechanics}, Addison-Wesley, 1967.
\bibitem{landau} See, for example, J. M. Blatt and V. F. Weisskopf,  
         {\it Theoretical Nuclear Physics}, John Wiley \& Sons, 1952; 
        	A. Messiah, 
                  {\it Quantum Mechanics}, John Wiley \& Sons, 1958. 
\bibitem{BCL} T. Barnes, F. E. Close, and H. J. Lipkin, 
              \Journal{\PRD}{68}{054006}{2003}.
\end{thebibliography}
\end{document}